\documentclass[fleqn,usenatbib]{mnras}

\usepackage{newtxtext,newtxmath}

\usepackage[T1]{fontenc}
\usepackage{ae,aecompl}

\usepackage{graphicx}	
\usepackage{amsmath}	

\newcommand{\rmf}{{\tt{RMF}}~}

\newcommand{\rprs}{$R_\mathrm{p}/R_\mathrm{s}$~}

\title[WASP-43b observed with \textit{TESS} and MuSCAT2]{Is the orbit of the exoplanet WASP-43b really decaying? \textit{TESS} and MuSCAT2 observations confirm no detection}

\author[Z. Garai et al.]{
Z. Garai,$^{1,2,3,20}$\thanks{E-mail: zgarai@gothard.hu}
T. Pribulla,$^{1,2,3}$
H. Parviainen,$^{4,5}$
E. Pall\'{e},$^{4,5}$
A. Claret,$^{6,7}$
L. Szigeti,$^{1,2,20}$
\newauthor
V. J. S. B\'{e}jar,$^{4,5}$
N. Casasayas-Barris,$^{19}$
N. Crouzet,$^{15}$ 
A. Fukui,$^{11,4}$
G. Chen,$^{17}$  
\newauthor
K. Kawauchi,$^{18}$
P. Klagyivik,$^{16}$
S. Kurita,$^{13}$
N. Kusakabe,$^{9,10}$
J. P. de Leon,$^{8}$
\newauthor
J. H. Livingston,$^{8}$
R. Luque,$^{6}$
M. Mori,$^{8}$
F. Murgas,$^{4,5}$
N. Narita,$^{4,9,11,12}$
\newauthor
T. Nishiumi,$^{9,14}$
M. Oshagh,$^{4,5}$
Gy. M. Szab\'{o}$^{1,2,20}$
M. Tamura,$^{8,9,10}$
\newauthor
Y. Terada,$^{8}$
N. Watanabe,$^{9,14}$
\\
$^{1}$MTA-ELTE Exoplanet Research Group, 9700 Szombathely, Szent Imre h. u. 112, Hungary\\
$^{2}$ELTE Gothard Astrophysical Observatory, 9700 Szombathely, Szent Imre h. u. 112, Hungary\\
$^{3}$Astronomical Institute, Slovak Academy of Sciences, 05960 Tatransk\'a Lomnica, Slovakia\\
$^{4}$Instituto de Astrof\'{i}sica de Canarias, 38200 La Laguna, Tenerife, Spain\\
$^{5}$Department Astrof\'{i}sica, Universidad de La Laguna, 38206 La Laguna, Tenerife, Spain\\
$^{6}$Instituto de Astrof\'{i}sica de Andaluc\'{i}a, CSIC, Apartado 3004, 18080 Granada, Spain\\
$^{7}$Dept. F\'{i}sica Te\'{o}rica y del Cosmos, Universidad de Granada, Campus de Fuentenueva s/n, 10871 Granada, Spain\\ 
$^{8}$Department of Astronomy, Graduate School of Science, The University of Tokyo, 7-3-1 Hongo, Bunkyo-ku, Tokyo 113-0033, Japan\\
$^{9}$Astrobiology Center, 2-21-1 Osawa, Mitaka-shi, Tokyo 181-8588, Japan\\
$^{10}$National Astronomical Observatory, 2-21-1 Osawa, Mitaka-shi, Tokyo 181-8588, Japan\\
$^{11}$Komaba Institute for Science, The University of Tokyo, 3-8-1 Komaba, Meguro, Tokyo 153-8902, Japan\\
$^{12}$JST, PRESTO, 3-8-1 Komaba, Meguro, Tokyo 153-8902, Japan\\
$^{13}$Department of Earth and Planetary Science, Graduate School of Science, The University of Tokyo, 7-3-1 Hongo, Bunkyo-ku, Tokyo 113-0033, Japan\\
$^{14}$Department of Astronomical Science, The Graduated University for Advanced Studies, SOKENDAI, 2-21-1, Osawa, Mitaka, Tokyo, 181-8588, Japan\\
$^{15}$European Space Agency (ESA), European Space Research and Technology Centre (ESTEC), Keplerlaan 1, 2201 AZ Noordwijk, The Netherlands\\
$^{16}$Institute of Planetary Research, German Aerospace Center, Rutherfordstrasse 2, 12489 Berlin, Germany\\
$^{17}$Key Laboratory of Planetary Sciences, Purple Mountain Observatory, Chinese Academy of Sciences, Nanjing 210023, PR China\\
$^{18}$Department of Multi-Disciplinary Sciences, Graduate School of Arts and Sciences, The University of Tokyo, 3-8-1 Komaba, Meguro, Tokyo 153-8902, Japan\\
$^{19}$Leiden Observatory, Leiden University, Postbus 9513, 2300 RA Leiden, The Netherlands\\
$^{20}$MTA-ELTE Lend\"{u}let Milky Way Research Group, Hungary 
}

\date{Accepted XXX. Received YYY; in original form ZZZ}

\pubyear{2015}

\begin{document}
\label{firstpage}
\pagerange{\pageref{firstpage}--\pageref{lastpage}}
\maketitle

\begin{abstract}
Up to now, WASP-12b is the only hot Jupiter confirmed to have a decaying orbit. The case of WASP-43b is still under debate. Recent studies preferred or ruled out the orbital decay scenario, but further precise transit timing observations are needed to definitively confirm or refute the period change of WASP-43b. This possibility is given by the \textit{Transiting Exoplanet Survey Satellite} (\textit{TESS}) space telescope. In this work we used the available \textit{TESS} data, multi-color photometry data obtained with the Multicolor Simultaneous Camera for studying Atmospheres of Transiting exoplanets 2 (MuSCAT2) and literature data to calculate the period change rate of WASP-43b and to improve its precision, and to refine the parameters of the WASP-43 planetary system. Based on the observed-minus-calculated data of 129 mid-transit times in total, covering a time baseline of about 10 years, we obtained an improved period change rate of $\dot{P} = -0.6 \pm 1.2$ ms~yr$^{-1}$ that is consistent with a constant period well within $1\sigma$. We conclude that new \textit{TESS} and MuSCAT2 observations confirm no detection of WASP-43b orbital decay.

\end{abstract}




\section{Introduction}
\label{intro}

Transit time variations (TTVs) of known exoplanets can be used to search for a third body in the planetary systems, but variations of transit times can also indicate star-planet tidal interaction. This interaction has various forms: \textit{apsidal precession} in which the orbit ellipse rotates in its own plane, and \textit{nodal precession} in which the orbit normal precesses about the total angular momentum vector. For eccentric orbits, both will result in long-term variations of the transit times, and of the transit durations. Typically, apsidal precession dominates \citep{Miralda1, Ragozzine1}. In addition to the effects of apsidal precession due to tidal bulges, tidal effects in close-in planets can lead to tidal decay, and a shift in transit times \citep{Sasselov1}. This tidal decay is also referred to as \textit{orbital decay} or tidal inspiraling. The end result of this process is tidal disruption and disintegration of the planet body.

Tidal decay was considered in several cases, for example, for WASP-18 \citep{Hellier2}, KELT-16 \citep{Oberst1}, WASP-103 \citep{Gillon2}, WASP-12 \citep{Hebb1} and WASP-43 \citep{Gillon1}. The full list of interesting planets from this viewpoint was summarized in Table 1 of \citet{Patra2}. In the case of WASP-4 a decreasing orbital period was detected by \citet{Bouma1}, later \citet{Bouma2} concluded that the system is accelerating toward the Sun, and the associated Doppler effect should cause the apparent period change rate. Up to now, WASP-12b is the only hot Jupiter to have a decaying orbit confirmed by \citet{Turner2}. The confirmation was a long process started by \citet{Maciejewski3}, who reported as first the possibility that the orbital period of WASP-12b is changing. Further observations confirmed the departure of transit times from the linear ephemeris \citep{Patra1, Maciejewski4, Bailey1}. \citet{Yee1} preferred orbital decay over apsidal precession or Romer effect. \citet{Turner2} analyzed, besides the literature data, the data obtained by \textit{TESS} \citep{Ricker1} to characterize the system and to verify that the planet is undergoing orbital decay. The authors highly favor the orbital decay scenario and obtained a decay rate of $\dot{P} = -32.5 \pm 1.6$ ms~yr$^{-1}$.

The case of WASP-43b is still under debate. The investigation of its possible orbital decay started with the orbital period change rate of $\dot{P} = -95 \pm 36$ ms~yr$^{-1}$, reported by \citet{Blecic1}. In the same year, \citet{Murgas1} and \citet{Chen1} revised the period change rate and obtained a value of $\dot{P} = -150 \pm 60$ ms~yr$^{-1}$ and $\dot{P} = -90 \pm 40$ ms~yr$^{-1}$, respectively. Further estimation of a period change rate of $\dot{P} = -30 \pm 30$ ms~yr$^{-1}$ was presented by \citet{Ricci1}. This result left open the question of the period change of WASP-43b. In the next year a possible orbital change rate of $ \dot{P} = -28.9 \pm 7.7$ ms~yr$^{-1}$ was presented and the orbital decay scenario was preferred by \citet{Jiang1}, on the other hand \citet{Hoyer1} ruled out the orbital decay based on additional transit observations, obtaining a period change rate of $\dot{P} = 0.0 \pm 6.6$ ms~yr$^{-1}$. \citet{Stevenson1} added three \textit{Spitzer Space Telescope} transit observations and found no evidence for orbital decay, moreover the obtained change rate was positive ($\dot{P} = +9.0 \pm 4.0$ ms~yr$^{-1}$). Similarly, \citet{Patra2} increased the time baseline by adding three transits and found that the period of WASP-43b has changed slightly. They found a period change rate of $\dot{P} = +14.4 \pm 4.6$ ms~yr$^{-1}$. The authors noted that this result could be only a statistical fluke, and this will be clearer after more observations in the future. Thus, the possible orbital decay of the exoplanet WASP-43b is still an open question, and precise transit timing observations are needed to definitively confirm or refute the orbital decay scenario. This possibility is given by the \textit{TESS} space telescope. \textit{TESS} provides high-precision photometric transit observations, which we used for searching for TTVs. 

In this paper we aimed at refining the system parameters based on \textit{TESS} and MuSCAT2 data and to calculate the period change rate of WASP-43b. We combined \textit{TESS} data and multi-color ground-based observations, because multi-color photometric transit observations can ameliorate the degeneracy between the planet-to-star radius ratio and the orbit inclination angle. The first parameter is passband dependent, but the second parameter is passband independent \citep{Csizmadia1, Espinoza1, Parviainen2}. The paper is organized as follows. In Section \ref{obs} a brief description of instrumentation and data reduction is given. Data analysis and model fitting are detailed in Section \ref{dataanalysis}. Our results are described in Section \ref{results}. We summarize our findings in Section \ref{concl}.                                    

\section{Observations and data reduction}
\label{obs}

WASP-43b was observed with \textit{TESS} in Sector No. 9, from 2019-02-28 to 2019-03-26 and also in Sector No. 35, from 2021-02-09 to 2021-03-07. The data were downloaded from the Mikulski Archive for Space Telescopes\footnote{See \url{https://mast.stsci.edu/portal/Mashup/Clients/Mast/Portal.html}.} in the form of Simple Aperture Photometry (SAP) fluxes. Sector No. 9 contains 15~602 data-points, Sector No. 35 13~661 data-points, i.e., we used 29~263 \textit{TESS} data-points in total during our analysis. The number of transits observed in Sector No. 9 is 26, while from Sector No. 35 we obtained 24 transits, i.e., we used 50 \textit{TESS} transits in total during the analysis (see Table \ref{TESSobslog} for the \textit{TESS} observational log). These data were obtained from 2-min integrations, but in comparison with Pre-search Data Conditioning Simple Aperture Photometry (PDCSAP) fluxes, long-term trends were not removed. The downloaded SAP fluxes were detrended using our pipeline, described later in this section.    

\begin{table}
\centering
\caption{Log of \textit{TESS} photometric observations of WASP-43 used in our analysis (sorted by \textit{TESS} sectors). Table shows time interval of observations, number of observed transits, and number of data points obtained from the \textit{TESS} database.}
\label{TESSobslog}
\begin{tabular}{cccc}
\hline
\hline
\textit{TESS} Sector & Time interval of observations & Transits & Data points\\
\hline
\hline
No. 09		& 2019-02-28 -- 2019-03-26	& 26		& 15~602\\
No. 35		& 2021-02-09 -- 2021-03-07	& 24		& 13~661\\
\hline
Total		& --				& 50		& 29~263\\
\hline
\hline
\end{tabular}
\end{table}

\begin{figure}
\includegraphics[width=\columnwidth]{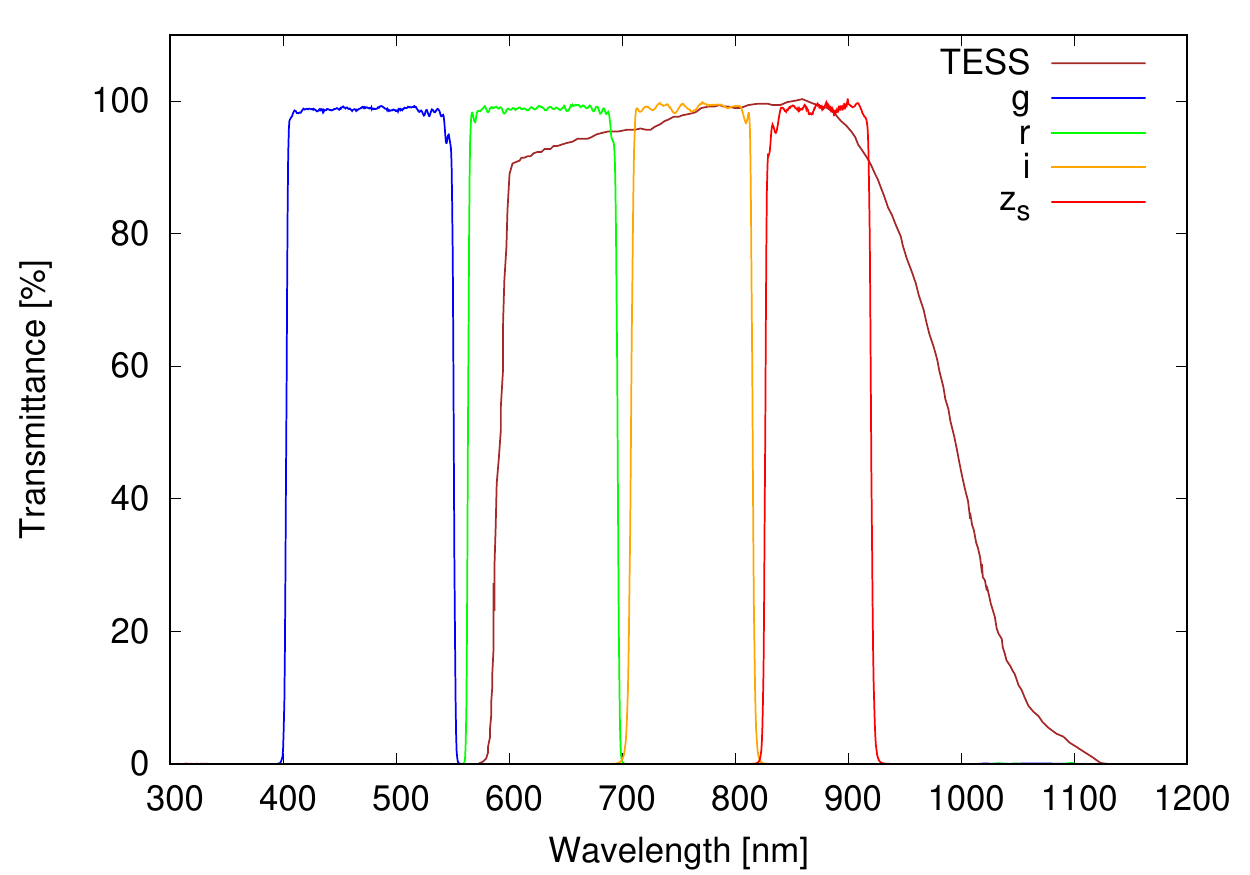}
\caption{Relative transmittance of the \textit{TESS} instrument and the modified Sloan $g$, $r$, $i$, and $z_s$ filters, installed on the MuSCAT2 instrument. The tabulated data were obtained from the web (\url{https://heasarc.gsfc.nasa.gov/docs/tess/the-tess-space-telescope.html}) and by personal communication from Norio Narita (\url{narita@g.ecc.u-tokyo.ac.jp}), respectively.}
\label{transmit}
\end{figure}

The multi-color photometric observations were performed using the Carlos S\'{a}nchez Telescope (TCS) on the island of Tenerife (Spain). The TCS is a 1.52 m diameter Cassegrain type $f/13.8$ telescope, installed on an equatorial mount\footnote{See \url{http://research.iac.es/OOCC/iac-managed-telescopes/telescopio-carlos-sanchez/}.}. The photometric detector, called MuSCAT2, is installed in the Cassegrain focus of the telescope. The number "2" means that this is already the second such instrument. MuSCAT2 is a four-color dichroic instrument, with four $1024 \times 1024$ CCD cameras, made by Princeton Instruments, having a field of view of $7.4' \times 7.4'$ with a pixel scale of $0.44''$ per pixel. Three dichroic mirrors separate incoming light into the four CCDs, enabling simultaneous operation of the cameras. No filter wheel is applied, every CCD has only one filter. Modified Sloan filters $g$, $r$, $i$, and $z_s$ are used with the transparencies from 400 to 550 nm for the $g$ filter, from 550 to 700 nm for the $r$ filter, from 700 to 820 nm for the $i$ filter, and from 820 to 920 nm for the $z_s$ filter, see Fig. \ref{transmit}. More details about this instrument can be found in \citet{Narita2}. 

The observations of WASP-43 were carried out with telescope defocusing. Mainly the CCD camera equipped with the $g$ filter was used for guiding the telescope. We used MuSCAT2 transit observations of WASP-43b from five nights in total. These observations were carried out before \textit{TESS} observations. The evening dates of the MuSCAT2 observations are the followings: 2018-01-09, 2018-01-18, 2018-02-18, 2018-04-03, and 2019-01-02. We obtained 3690, 4877, 4306, and 3182 data-points in the passbands $g$, $r$, $i$, and $z_s$, respectively (see Table \ref{photobslog} for the MuSCAT2 observational log). The multi-color photometric observations were reduced using the dedicated MuSCAT2 photometry pipeline following \citet{Parviainen2}. The pipeline works under the {\tt{Python3}}\footnote{See \url{https://www.python.org/download/releases/3.0/}.} environment and it is based on {\tt{NumPy}} \citep{vanderWalt1}, {\tt{SciPy}} \citep{Virtanen1}, {\tt{AstroPy}} \citep{Astropy1}, and {\tt{Photutils}} \citep{Bradley1} packages. During the first step the pipeline makes dark and flat corrections of the scientific frames. After this step the astrometric solution is performed using the {\tt{Astrometry.net}} software \citep{Lang1}. Finally, the pipeline calculates aperture photometry. During this step the target star and up to 14 comparison stars in 10 apertures are calculated, which means that up to 150 absolute light curves are created per measurement and filter. During the next step we first identified the target star per measurement and filter, subsequently other stars were used as comparison stars and the relative light curve of the target star was calculated. We tried every comparison star and every aperture. In all cases the scatter of the corresponding relative light curve was determined and the best three comparison stars with the best apertures, which produced the lowest scatter of the corresponding relative light curve of the target star, were selected. As the final step we prepared the relative transit light curve per measurement and filter by using the best three comparison stars with the best apertures as an average comparison star.            

\begin{table}
\centering
\caption{Log of MuSCAT2 multi-color photometric observations of WASP-43 used in our analysis (sorted by evening dates). Table shows number of scientific frames per passband ($N_\mathrm{passband}$), obtained during the given observing night and the applied exposure times in seconds ($Exp_\mathrm{passband}$).}
\label{photobslog}
\begin{tabular}{ccccc}
\hline
\hline
Evening date & $N_\mathrm{g}/Exp_\mathrm{g}$ & $N_\mathrm{r}/Exp_\mathrm{r}$ & $N_\mathrm{i}/Exp_\mathrm{i}$ & $N_\mathrm{z_s}/Exp_\mathrm{z_s}$\\
\hline
\hline
2018-01-09	    &	2086/5			& 2078/5	    & 2076/5		    & 1425/8\\
2018-01-18          &   835/13         		& 835/13            & 835/13                & 835/13\\
2018-02-18	    &	328/30		    	& 898/10	    & 626/15		    & 481/20\\
2018-04-03	    &	441/9		    	& 441/9		    & 441/9		    & 441/9\\
2019-01-02	    &	--/--		    	& 625/15	    & 328/30		    & --/--\\
\hline
Total		    &   3690/--                 & 4877/--           & 4306/--               & 3182/--\\
\hline
\hline
\end{tabular}
\end{table}

The MuSCAT2 multi-color light curves were first normalized to unity. After that, the linear trend, due to the second-order extinction, was removed from the photometric data. As a next step we cleaned the light curves from outliers. We used a $3\sigma$ clipping, where $\sigma$ is the standard deviation of the light curve. Subsequently, we converted all remaining time-stamps from Modified Julian Date in Universal Time Coordinated ($\mathrm{MJD}_\mathrm{UTC}$), which is the used output-time-stamp format of the MuSCAT2 photometry pipeline (i.e., $\mathrm{JD} - 2~400~000.5$), to Barycentric Julian Date in Barycentric Dynamical Time ($\mathrm{BJD}_\mathrm{TDB}$), using the online applet {\tt{UTC2BJD}}\footnote{See \url{http://astroutils.astronomy.ohio-state.edu/time/utc2bjd.html}.} \citep{Eastman1}.

\begin{figure}
\centering
\includegraphics[width=\columnwidth]{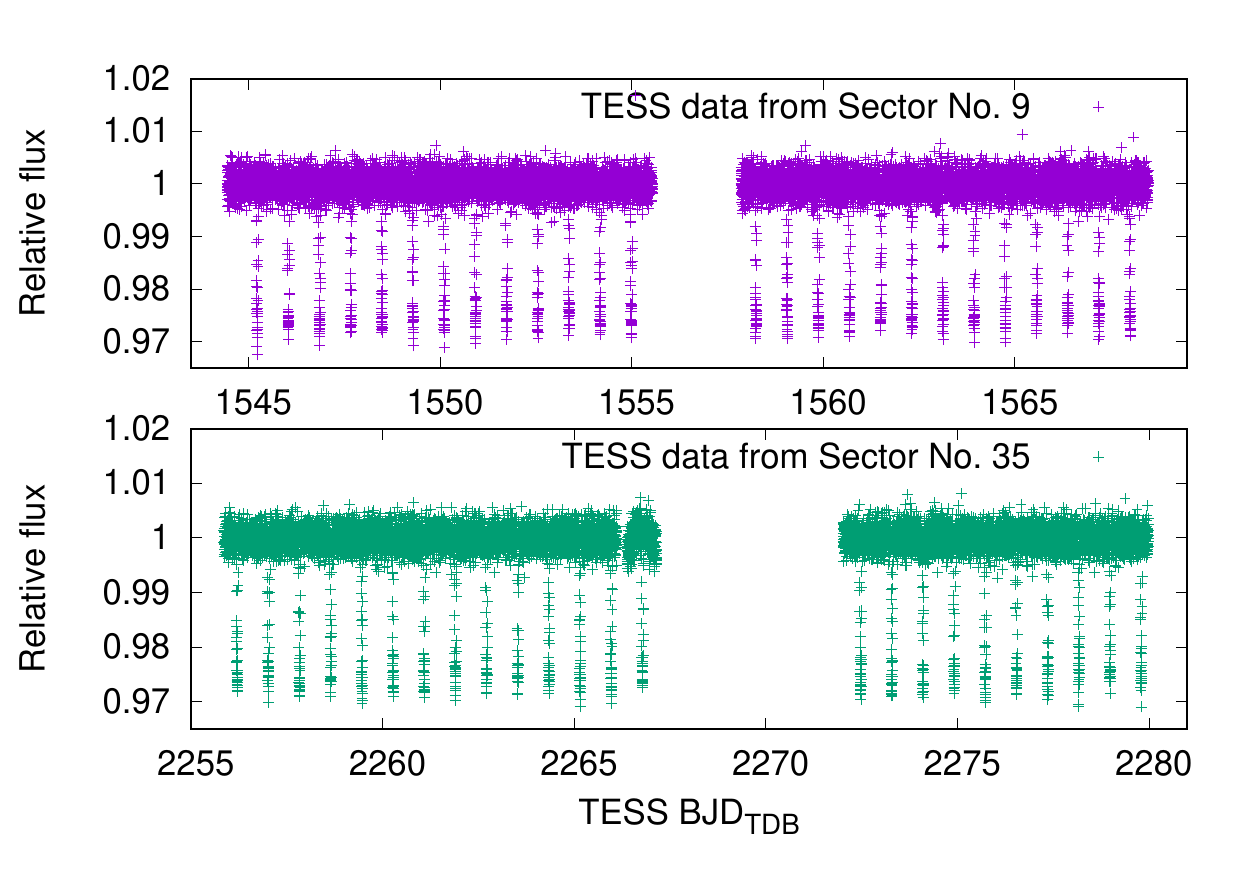}
\caption{Detrended and normalized \textit{TESS} data taken from the Mikulski Archive for Space Telescopes in the form of SAP fluxes.}
\label{tessdata} 
\end{figure}

The downloaded \textit{TESS} data were treated similarly as MuSCAT2 data. The SAP fluxes were first normalized to unity. During the next step \textit{TESS} data were cut into segments, each covering one orbital period. Each segment of the data was fitted with a linear function. During the fitting procedure the part of the data covering the transit was excluded from the fit. Consequently, the linear trend was removed from each chunk of data (including the transit data). This detrending method can effectively remove the long-term variability (mainly variability of the host star due to spots and rotation) while it does not introduce any nonlinear trend to the data, see Fig. \ref{tessdata} and, e.g., \citet{Garai2}. Outliers were cleaned similarly as in the case of the MuSCAT2 observations. Since \textit{TESS} uses as time-stamps Barycentric \textit{TESS} Julian Date (i.e., $\mathrm{BJD}_\mathrm{TDB} - 2~457~000.0$), during the next step we converted all \textit{TESS} time-stamps to $\mathrm{BJD}_\mathrm{TDB}$. 

During the modeling tests (See Sect. \ref{transitmodel}) we recognized that mainly in the case of MuSCAT2 data the linear detrending is not enough, also correlated noise is present in the data. To better detrend the data and to remove correlated noise we applied the following procedure. We first detrended the MuSCAT2 data using higher-order polynomial, then we fitted every single light curve, including \textit{TESS} transits, using the Levenberg-Marquardt method implemented in the non-linear least-squares minimization and light-curve fitting package, called {\tt{lmfit}}\footnote{See \url{https://lmfit.github.io/lmfit-py/}.}. The following free parameters and priors were adjusted: the mid-transit time $T_\mathrm{c}$ = $N$(2~455~528.868634, 0.000046) $\mathrm{BJD}_\mathrm{TDB}$ \citep{Hoyer1}, the planet-to-star area ratio $(R_\mathrm{p}/R_\mathrm{s})^2$ = $N$(0.026, 0.001) calculated based on \citet{Hoyer1}, the transit duration $T_\mathrm{dur}$ = $N$(0.061, 0.001) in units of phase \citep{Esposito1}, the impact parameter $b$ = $N$(0.689, 0.013) in units of stellar radius \citep{Esposito1}, and the light-curve normalization factor $l_\mathrm{norm}$ = $U$(1.0, 0.5) in fluxes. The quadratic limb darkening coefficients were calculated for the MuSCAT2 $g$, $r$, $i$, and $z_\mathrm{s}$ passbands based on the spherical {\tt{PHOENIX-COND}} models \citep{Claret1} and then linearly interpolated using the stellar parameters of $T_\mathrm{eff} = 4500 \pm 100$ K, log $g = 4.5 \pm 0.2$ cgs, and Fe/H = $-0.01 \pm 0.15$ dex, see \citet{Esposito1}. For the \textit{TESS} passband the quadratic limb-darkening coefficients were linearly interpolated from the Table 5\footnote{See \url{https://cdsarc.unistra.fr/viz-bin/nph-Cat/html?J/A+A/618/A20/table5.dat}.}, published by \citet{Claret1}. These coefficients were fixed during the fitting procedure (see Sect. \ref{transitmodel}), as well as the orbital period of WASP-43b adopted from \citet{Hoyer1}, i.e., $P = 0.813473978$ d. We assumed circular orbit of WASP-43b.

\begin{figure}
\centering
\includegraphics[width=\columnwidth]{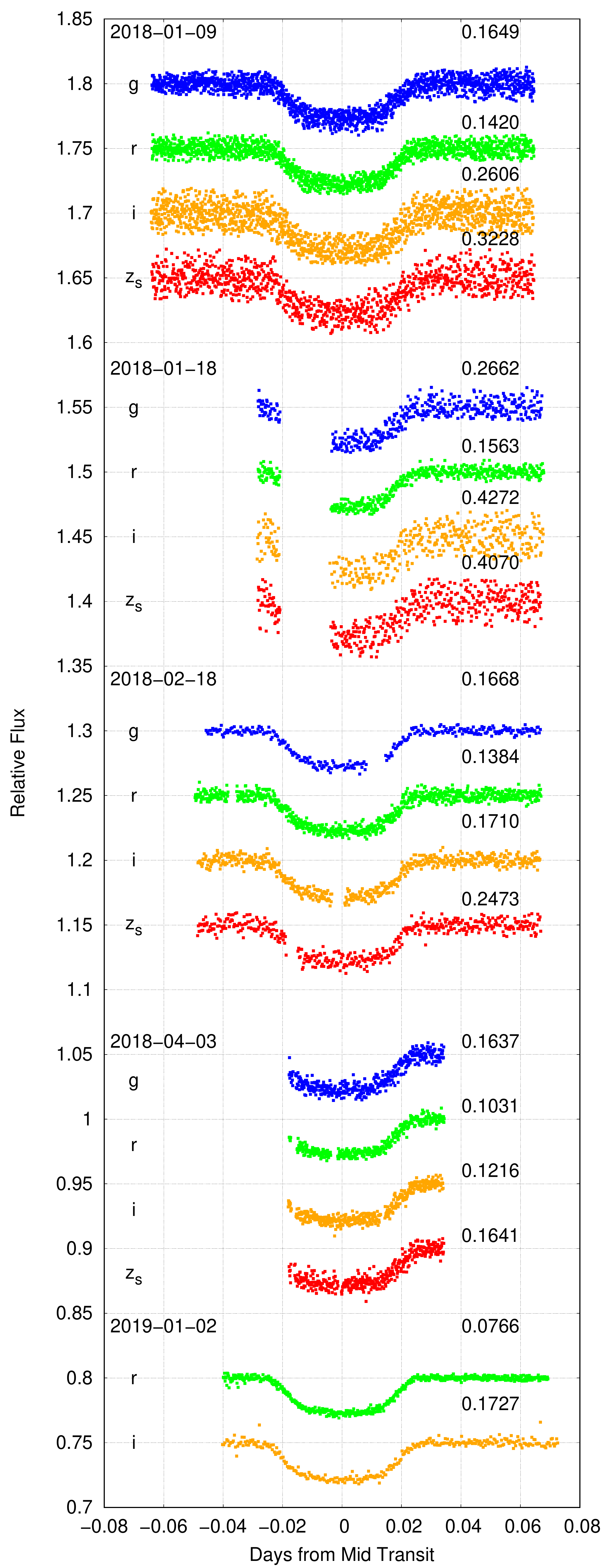}
\caption{Detrended MuSCAT2 data, plotted in chronological order starting from the top, and offset in relative flux for clarity. On the right side of the graph the quality of individual light curves is quantified using $PNR$ (in \%~min$^{-1}$) for comparison purposes (see the last paragraph of Sect. \ref{obs}).}
\label{indlcmuscat} 
\end{figure}

As a next step we ran a Markov chain Monte Carlo (MCMC) analysis with the affine-invariant sampler implemented in the {\tt{emcee}}\footnote{See \url{https://emcee.readthedocs.io/en/stable/}.} package \citep{Foreman1}. During this step we also modeled correlated noise using a Gaussian process regression method with the {\tt{SHOTerm}} plus {\tt{JitterTerm}} kernel, with a fixed quality factor $Q = 1/\sqrt{2}$, implemented in the {\tt{CELERITE}}\footnote{See \url{https://celerite.readthedocs.io/en/stable/}.} package \citep{Kallinger1, Foreman2, Barros1}. The regression is done by using $\log \sigma$ (free), $\log Q$ (fixed), $\log \omega_0$ (free), and $\log S_0$ (free) hyperparameters with bounds on the values of these parameters to be inputted by the user. We first fixed the transit shape, i.e., the parameters $(R_\mathrm{p}/R_\mathrm{s})^2$, $T_\mathrm{dur}$, $b$, and the mid-transit time $T_\mathrm{c}$ from the {\tt{lmfit}} result and set free the three hyperparameters for a preliminary MCMC analysis. The posteriors of the hyperparameters obtained from this analysis were used to define the priors for the next MCMC analysis as twice the uncertainty computed from the posterior distribution. Finally, we ran the MCMC analysis again with free transit model parameters and free hyperparameters. As the very last step we removed correlated noise from the data. We can note that \textit{TESS} data changed negligibly with this procedure, however MuSCAT2 data were significantly detrended. These detrended MuSCAT2 light curves of WASP-43b transits are depicted in Fig. \ref{indlcmuscat}.

To quantify and compare the quality of individual detrended light curves we used a quantity, which is called photometric noise rate ($PNR$), adopted from \citet{Fulton1}. It is defined as:       

\begin{equation}
\label{}
PNR = \frac{rms}{\sqrt{\Gamma}},
\end{equation} 

\noindent
where the root mean square ($rms$) is derived from the light curve residuals and $\Gamma$ is the median number of cycles, including exposure time and readout time\footnote{Readout time of MuSCAT2 CCD cameras is about 1.35 s, while in the case of \textit{TESS} CCD camera it is about 0.3 s.}, per minute. MuSCAT2 $PNR$ values are summarized in Fig. \ref{indlcmuscat} and \textit{TESS} transit light curves are qualified with $PNR$ in Table \ref{wasp43times}. We can see that the MuSCAT2 photometry in $z_\mathrm{s}$ passband has always the worst quality, the best quality measurement was taken on 2019-01-02 in $r$ passband. \textit{TESS} light curves have stable quality with $PNR \approx 0.26$ \%~min$^{-1}$, which is comparable to the MuSCAT2 measurement taken on 2018-01-18 in $g$ passband.

\section{Data analysis}
\label{dataanalysis}

\subsection{Transit modelling}
\label{transitmodel}

We analyzed the detrended \textit{TESS} and MuSCAT2 photometry data using the \rmf (Roche ModiFied) code. The software was prepared only recently based on the {\tt{ROCHE}} code, which is devoted to the modeling of multi-data set observations of close eclipsing binary stars, such as radial velocities and multi-color light curves \citep{Pribulla2}. The \rmf code was already used with success, e.g., in \citet{Szabo1}. The code can simultaneously model multi-color light curves, radial velocities, and broadening functions, or least-squares deconvolved line profiles of binary stars and transiting exoplanets. Its modification to be used with the transiting exoplanets uses the Roche surface geometry with the planet gravity neglected for the host star (rotationally deformed shape) and spherical shape for the planet. The model can handle eccentric orbits, misaligned rotational axes of the components, stellar oblateness, gravity darkening due to rapid rotation using the analytical approach of \citet{Espinosa1}, Doppler beaming effect, advanced limb-darkening description, and third light. The synthesis of the broadening functions assumes solid-body rotation. The synthesis of the observables is performed in the plane of the sky using pixel elements. The effectiveness of the integration is increased by the adaptive phase step being more fine during the eclipses/transits.            

During the analysis procedure we used the following parameters: the mid-transit time $T_\mathrm{c}$, the orbital period $P_\mathrm{orb}$, the orbit inclination angle $i$ with respect to the plane of the sky, the ratio of the host radius to the semi-major axis $R_\mathrm{s}/a$, the passband-dependent planet-to-star radius ratio $R_\mathrm{p}/R_\mathrm{s}$, the eccentricity $e$, the longitude of the periastron passage $\omega$, the passband-dependent third light $l_3$, defined as $l_3/(l_1+l_2)$, and the passband-dependent light-curve normalization factor $l_\mathrm{norm}$. The stellar limb darkening is described by the four-parameter limb-darkening model of \citet{Claret1} with the critical foreshortening angle when the intensity drops to zero\footnote{Parameter $\mu = \cos \theta$, where $\theta$ is the so-called foreshortening angle, which is angle between the line of sight and a normal to the stellar surface. For $\mu < \mu_\mathrm{crit}$ the stellar flux is assumed to be zero, see \citet{Claret1}.}, interpolating for local gravity and temperature for each pixel. The limb-darkening coefficients ($a_1$, $a_2$, $a_3$, $a_4$, and $\mu_\mathrm{crit}$) were calculated for the $g$, $r$, $i$, $z_\mathrm{s}$, and \textit{TESS} passbands, using the tabulated transmittance of the MuSCAT2 filters and \textit{TESS} instrument, and using the same spherical {\tt{PHOENIX-COND}} models as in \citet{Claret1}. The limb-darkening coefficients were linearly interpolated from the calculated tables of coefficients for the stellar parameters of $T_\mathrm{eff} = 4500 \pm 100$ K, log $g = 4.5 \pm 0.2$ cgs, and Fe/H = $-0.01 \pm 0.15$ dex, see \citet{Esposito1}. These coefficients were fixed during the fitting procedure. We note that prior this treatment we ran several test modelings with quadratic and four-parameter limb darkening coefficients allowed to float. Since we always got unphysical fitted coefficients far from the tabulated theoretical values, we decided to keep fixed these coefficients during the fitting procedure and to use the four-parameter model, because there is no significant difference in fitted parameters while applying either the quadratic or the four parameter approach, but the latter one represents better the distribution of specific intensities, because it improves the description of both the stellar limb and the central parts. The unphysical fitted limb-darkening coefficients could be due to the high impact parameter of the system, $b= 0.689 \pm 0.013$ \citep{Esposito1}, which means that the transit chord is located in such a disk region, where $\mu$ values are from a very narrow interval. In addition, the eccentricity $e$ was set to zero and the longitude of the periastron passage $\omega$ was fixed at $90^{\circ}$, i.e., we assumed circular orbit of WASP-43b. Finally, we also fix the $l_3$ parameters. In the case of MuSCAT2 apertures there is no third light contamination. \textit{TESS} has larger aperture than MuSCAT2, therefore the third light contamination possibility is also larger. Since we used SAP fluxes, which are not corrected by the dilution factor, we used the CROWDSAP\footnote{CROWDSAP is a keyword on the header of the fits files containing the light curves. It represents the ratio of the target flux to the total flux in the \textit{TESS} aperture.} crowding metric value to determine the $l_3$ parameter for the \textit{TESS} aperture. In the case of WASP-43b this gives $l_{3,\mathrm{\textit{TESS}}} = 0.0008$.        

The joint model optimization was done by the steepest descend method, using the numerical derivatives of the observables with respect to the parameters. The optimization was run until the $\chi^2$ improvement was smaller than 0.00005. To obtain realistic estimates of the parameter uncertainties 2000 Monte Carlo experiments were performed. The artificial data-sets were created from the best fitting model at the times of observations adding a random Gaussian noise equal to the standard deviation of the data with respect to the fit. We aimed at compensating the fixed limb-darkening coefficients, therefore uncertainty of the parent star's effective temperature and surface gravity, which strongly affect limb darkening, were propagated to the modeling as Gaussian priors. The distribution of the parameters were analyzed to obtain the final values and the standard errors. The best-fitting parameter values correspond to quantile 0.50 (median) and the uncertainties to quantils $\pm 0.341$.

\begin{table}
\centering
\caption{The list of the observed ('O') mid-transit times of WASP-43b, derived using the joint model parameter values (see Table \ref{wasp43outputs}). During this fitting procedure only the mid-transit time and the light-curve normalization factor parameters were adjusted. The uncertainties in the fitted parameters were estimated based on the covariance matrix method. The quality of individual \textit{TESS} light curves is also quantified using $PNR$ for comparison purposes (see the last paragraph of Sect. \ref{obs}).}
\label{wasp43times}
\begin{tabular}{cccc}
\hline
\hline
Transit & 'O' times [$\mathrm{BJD}_\mathrm{TDB}$] & $PNR$ [\%~min$^{-1}$] & Source\\
\hline
\hline
No. 01 & $2458128.732027 \pm 0.000080$ & See Fig. \ref{indlcmuscat} & MuSCAT2\\
No. 02 & $2458137.680280 \pm 0.000120$ & See Fig. \ref{indlcmuscat} & MuSCAT2\\
No. 03 & $2458168.592124 \pm 0.000068$ & See Fig. \ref{indlcmuscat} & MuSCAT2\\
No. 04 & $2458212.519760 \pm 0.000070$ & See Fig. \ref{indlcmuscat} & MuSCAT2\\
No. 05 & $2458486.660595 \pm 0.000064$ & See Fig. \ref{indlcmuscat} & MuSCAT2\\
No. 06 & $2458545.23075 \pm 0.00020$ & 0.2710 & \textit{TESS}\\
No. 07 & $2458546.04416 \pm 0.00022$ & 0.2620 & \textit{TESS}\\
No. 08 & $2458546.85750 \pm 0.00020$ & 0.2749 & \textit{TESS}\\
No. 09 & $2458547.67129 \pm 0.00021$ & 0.2826 & \textit{TESS}\\
No. 10 & $2458548.48448 \pm 0.00022$ & 0.2648 & \textit{TESS}\\
No. 11 & $2458549.29790 \pm 0.00020$ & 0.2884 & \textit{TESS}\\
No. 12 & $2458550.11120 \pm 0.00020$ & 0.2988 & \textit{TESS}\\
No. 13 & $2458550.92453 \pm 0.00021$ & 0.2684 & \textit{TESS}\\
No. 14 & $2458551.73864 \pm 0.00022$ & 0.2494 & \textit{TESS}\\
No. 15 & $2458552.55170 \pm 0.00021$ & 0.2683 & \textit{TESS}\\
No. 16 & $2458553.36520 \pm 0.00020$ & 0.2694 & \textit{TESS}\\
No. 17 & $2458554.17893 \pm 0.00021$ & 0.2529 & \textit{TESS}\\
No. 18 & $2458554.99224 \pm 0.00020$ & 0.2757 & \textit{TESS}\\
No. 19 & $2458558.24619 \pm 0.00021$ & 0.2767 & \textit{TESS}\\ 
No. 20 & $2458559.05960 \pm 0.00020$ & 0.2652 & \textit{TESS}\\
No. 21 & $2458559.87291 \pm 0.00020$ & 0.3016 & \textit{TESS}\\
No. 22 & $2458560.68623 \pm 0.00021$ & 0.2763 & \textit{TESS}\\
No. 23 & $2458561.49990 \pm 0.00020$ & 0.2584 & \textit{TESS}\\
No. 24 & $2458562.31369 \pm 0.00020$ & 0.2547 & \textit{TESS}\\
No. 25 & $2458563.12719 \pm 0.00022$ & 0.2864 & \textit{TESS}\\
No. 26 & $2458563.94020 \pm 0.00020$ & 0.2710 & \textit{TESS}\\
No. 27 & $2458564.75415 \pm 0.00021$ & 0.2727 & \textit{TESS}\\
No. 28 & $2458565.56763 \pm 0.00020$ & 0.2796 & \textit{TESS}\\
No. 29 & $2458566.38034 \pm 0.00020$ & 0.2949 & \textit{TESS}\\
No. 30 & $2458567.19410 \pm 0.00020$ & 0.2866 & \textit{TESS}\\
No. 31 & $2458568.00806 \pm 0.00020$ & 0.2798 & \textit{TESS}\\
No. 32 & $2459256.20649 \pm 0.00020$ & 0.2494 & \textit{TESS}\\
No. 33 & $2459257.02070 \pm 0.00021$ & 0.2538 & \textit{TESS}\\
No. 34 & $2459257.83368 \pm 0.00021$ & 0.2450 & \textit{TESS}\\
No. 35 & $2459258.64749 \pm 0.00020$ & 0.2502 & \textit{TESS}\\
No. 36 & $2459259.46044 \pm 0.00021$ & 0.2690 & \textit{TESS}\\
No. 37 & $2459260.27437 \pm 0.00021$ & 0.2585 & \textit{TESS}\\
No. 38 & $2459261.08729 \pm 0.00021$ & 0.2620 & \textit{TESS}\\
No. 39 & $2459261.90120 \pm 0.00021$ & 0.2618 & \textit{TESS}\\
No. 40 & $2459262.71463 \pm 0.00022$ & 0.2481 & \textit{TESS}\\
No. 41 & $2459263.52811 \pm 0.00020$ & 0.2549 & \textit{TESS}\\
No. 42 & $2459264.34187 \pm 0.00020$ & 0.2441 & \textit{TESS}\\
No. 43 & $2459265.15498 \pm 0.00021$ & 0.2630 & \textit{TESS}\\
No. 44 & $2459265.96887 \pm 0.00022$ & 0.2696 & \textit{TESS}\\
No. 45 & $2459266.78191 \pm 0.00021$ & 0.3170 & \textit{TESS}\\
No. 46 & $2459272.47657 \pm 0.00021$ & 0.2544 & \textit{TESS}\\
No. 47 & $2459273.28930 \pm 0.00020$ & 0.2695 & \textit{TESS}\\
No. 48 & $2459274.10301 \pm 0.00020$ & 0.2851 & \textit{TESS}\\
No. 49 & $2459274.91692 \pm 0.00020$ & 0.2735 & \textit{TESS}\\
No. 50 & $2459275.73038 \pm 0.00020$ & 0.2596 & \textit{TESS}\\
No. 51 & $2459276.54350 \pm 0.00020$ & 0.2639 & \textit{TESS}\\
No. 52 & $2459277.35728 \pm 0.00020$ & 0.2694 & \textit{TESS}\\
No. 53 & $2459278.17090 \pm 0.00021$ & 0.2469 & \textit{TESS}\\
No. 54 & $2459278.98399 \pm 0.00022$ & 0.2713 & \textit{TESS}\\
No. 55 & $2459279.79781 \pm 0.00020$ & 0.2644 & \textit{TESS}\\
\hline
\hline
\end{tabular}
\end{table}

\subsection{Timing analysis}
\label{tian}

To analyze the possible shift in transit times, which can indicate the orbital decay of WASP-43b, we constructed the so-called observed-minus-calculated (O-C) diagram for mid-transit times. We used 50 \textit{TESS} transits, 5 MuSCAT2 transits and literature data, which amount to 74 additional transits. To obtain the 'O' times of the mid-transits we modeled each \textit{TESS} and MuSCAT2 transit event individually using the \rmf code. During this procedure we fixed every parameter to its best value from the joint model except for two parameters: the mid-transit time and the light-curve normalization factor. The list of the fitted mid-transit times obtained from this modeling is presented in Table \ref{wasp43times}. The uncertainties in the fitted parameters were estimated based on the covariance matrix method. The literature data were taken directly in the form of mid-transit times. We used 68 mid-transit times compiled by \citet{Hoyer1}, see references therein, 3 mid-transit times obtained by \citet{Stevenson1}, and  3 mid-transit times of WASP-43b observed by \citet{Patra2}. We used 129 transit events in total, covering a time baseline of about 10 years. The first mid-transit time in our data-set, $\mathrm{BJD}_{\mathrm{TDB}}$ = 2~455~528.86863, corresponds to 08:50:49.63 UT on 2010-11-28. To calculate the 'C' times of the mid-transits we used the linear ephemeris formula as:    

\begin{equation}
\label{lineph}
T_0 = T_\mathrm{c} + P_\mathrm{orb} \times E,
\end{equation} 

\noindent were $T_0$ corresponds to the 'C' value, $T_\mathrm{c}$ is the reference mid-transit time, $P_\mathrm{orb}$ is the orbital period, and $E$ is the epoch of observation, i.e., the number of the orbital cycle calculated from $T_\mathrm{c}$. We used the jointly fitted parameter values for $T_\mathrm{c}$ and $P_\mathrm{orb}$ to calculate $T_0$ values (see Sect. \ref{sysparams}). In such a way we could construct the O-C diagram of WASP-43b mid-transit times. For this purpose we used the environment of the {\tt{OCFIT}}\footnote{See \url{https://github.com/pavolgaj/OCFit}.} code \citep{Gajdos2}. The software is simple thanks to a very intuitive graphic user interface. As first, we fitted the O-C data with a linear function using the {\tt{OCFIT}} package {\tt{FitLinear}}. The free parameters of the linear model are the reference mid-transit time $T_\mathrm{c}$ and the orbital period $P_\mathrm{orb}$. Subsequently, the O-C data were fitted with a quadratic function, also offered by the {\tt{OCFIT}} code, within the package called {\tt{FitQuad}}. The free parameters of the quadratic model are the reference mid-transit time $T_\mathrm{c}$, the orbital period $P_\mathrm{orb}$, and the quadratic coefficient $Q$, which follows from the quadratic ephemeris formula of:

\begin{equation}
\label{}
T_0 = T_\mathrm{c} + P_\mathrm{orb} \times E + Q \times E^2,
\end{equation}          

\noindent where the quadratic coefficient $Q$ can be expressed as:

\begin{equation}
\label{qqq}
Q = \frac{1}{2}P_\mathrm{orb} \times \dot{P},
\end{equation}          

\noindent where $\dot{P}$ means the orbital period change with time $t$, i.e., this is the so-called orbital period change rate: $\dot{P} = \mathrm{d}P/\mathrm{d}t$. $\dot{P}$ is a dimensionless quantity, but it can be expressed in s~yr$^{-1}$ or in ms~yr$^{-1}$. The uncertainties in the fitted parameters of $P_\mathrm{orb}$, $T_\mathrm{c}$, and $Q$ were derived within the {\tt{OCFIT}} packages {\tt{FitLinear}} and {\tt{FitQuad}} applying the covariance matrix method.

\section{Results}
\label{results}

\subsection{System parameters from the \textit{TESS} and MuSCAT2 data}
\label{sysparams}

\begin{figure}
\centering
\includegraphics[width=\columnwidth]{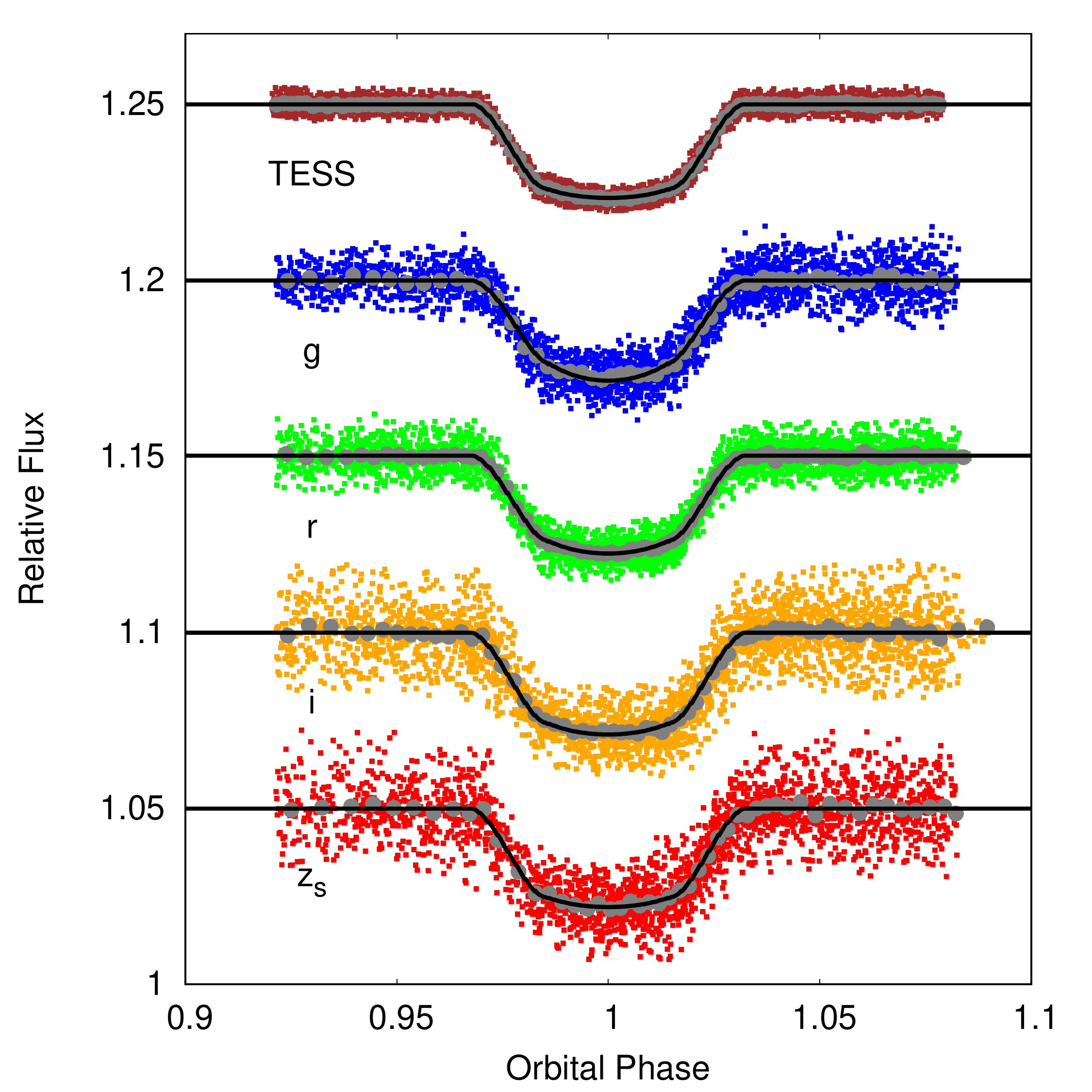}
\caption{Phase-folded transit light curves of WASP-43b in \textit{TESS} and MuSCAT2 $g$, $r$, $i$, and $z_s$ passbands, overplotted with the best-fitting \rmf models (black lines). We binned the data only for better visualization of the transit shape, but we fitted individual data-points (1 gray bin-point represents 50 data-points). During the joint modeling procedure all individual light curves per filter were combined and fitted simultaneously.}
\label{rmflcs}
\end{figure}

\begin{figure*}
\centering
\includegraphics[width=180mm]{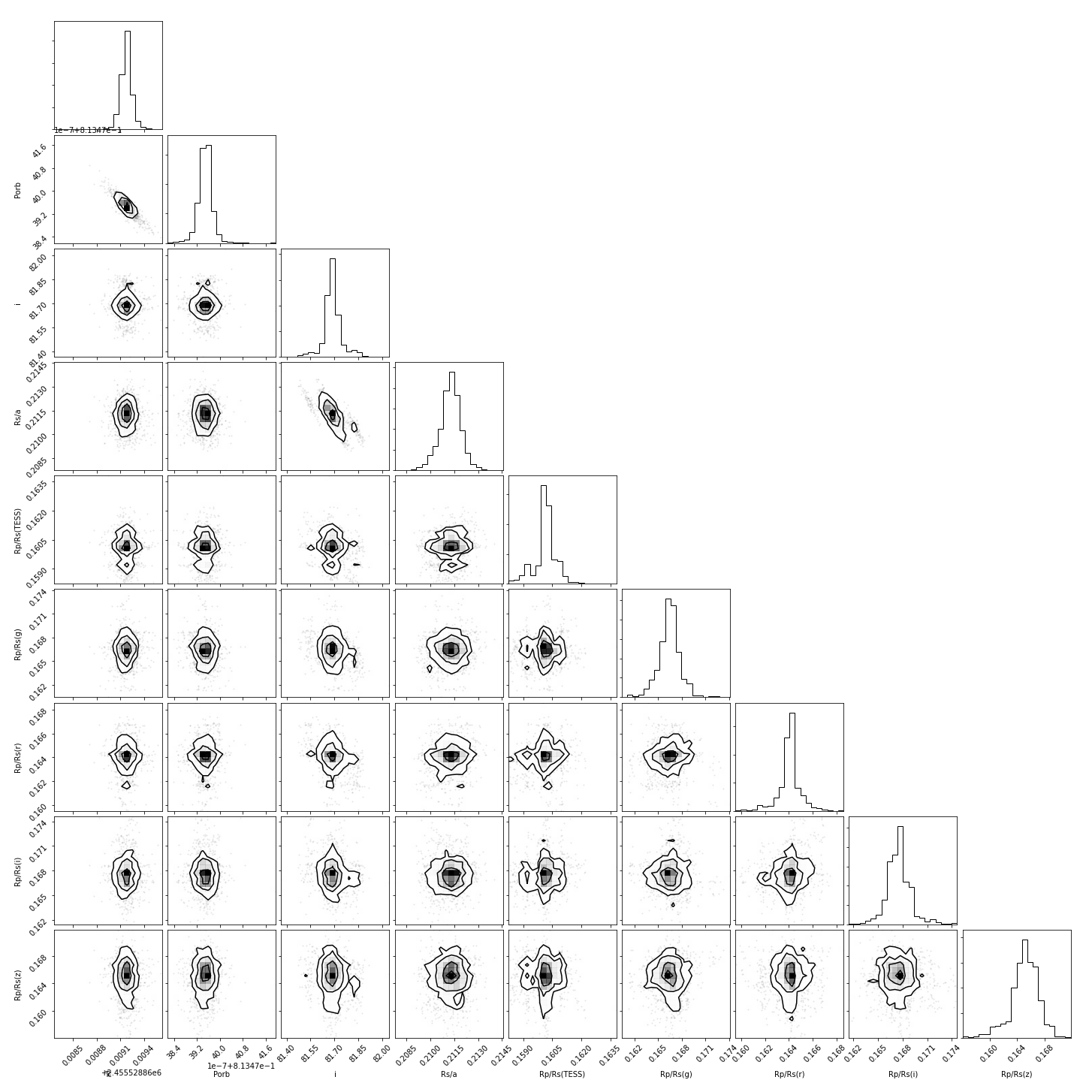}
\caption{Posterior probability distributions of WASP-43b, obtained from 2000 Monte Carlo experiments. The diagonal panels show the 1D distributions and the other panels show the 2D distributions and illustrate the parameter degeneracies.}
\label{corner} 
\end{figure*}

We summarize the fitted and derived parameters of the planetary system in Table \ref{wasp43outputs}. The phase-folded and binned transit light curves of the exoplanet WASP-43b, overplotted with the best-fitting \rmf models are presented in Fig. \ref{rmflcs}. The posterior probability distributions are depicted in Fig. \ref{corner}. WASP-43b is a dense gaseous planet despite its close-in orbit, which is due to the low effective temperature of the host star ($T_\mathrm{eff} = 4403^{+46}_{-53}$ K), and, consequently, to the relatively low equilibrium temperature of the planet ($T_\mathrm{eq} = 1426.7 \pm 8.5$ K, see \citet{Esposito1}). The radius of WASP-43b is very close to the radius of the planet Jupiter, i.e., $R_\mathrm{p} = 1.037^{+0.022}_{-0.019}$ $\mathrm{R}_\mathrm{Jup}$, but the mass of the planet is about 2-times the mass of Jupiter, i.e., $M_\mathrm{p} = 1.997^{+0.072}_{-0.073}$ $\mathrm{M}_\mathrm{Jup}$. The combination of these two parameters results in a planet density, which is about 1.6-times the density of the planet Jupiter ($\rho_\mathrm{p} = 2.10^{+0.31}_{-0.32}$ $\mathrm{g.cm}^{-3}$). WASP-43b had the closest orbit to its host star among hot Jupiters at the time of its discovery. The semi-major axis of the planet is $a = 0.01507^{+0.00026}_{-0.00027}$ a.u. 

\begin{table}
\centering
\caption{An overview of the best-fitting and derived parameters of the WASP-43 planetary system (host and planet b), obtained from the \textit{TESS} and MuSCAT2 data. The final fitted values correspond to quantile 0.50 (median) and the uncertainties to quantils $\pm 0.341$ in the parameter distributions obtained from 2000 Monte Carlo experiments. The values of $T_\mathrm{eff}$ and log $g$ are distributions of Gaussian priors for limb darkening. The $T_\mathrm{c}$ and $P_\mathrm{orb}$ parameter values are preliminary, see Sect. \ref{trantimings} and Table \ref{wasp43eph} for improved values. The planet-to-star radius ratio parameter value for all passbands combined was calculated as weighted average of \textit{TESS} and MuSCAT2 \rprs values with weights of $1/\sigma^{2}$, where $\sigma$ is the uncertainty in each passband. Notes: \textit{Gaia} DR2 = \citet{Gaia1}, E2017 = \citet{Esposito1}.}
\label{wasp43outputs}
\begin{tabular}{llc}
\hline
\hline
Parameter							& Value						& Source\\
\hline
\hline
\textit{Gaia} ID						& 3767805209112436736				& \textit{Gaia} DR2\\
RA [h:m:s] (J2000) 						& 10:19:38.0					& \textit{Gaia} DR2\\
Dec [deg:m:s] (J2000)						& -09:48:22.6					& \textit{Gaia} DR2\\
Parallax [mas]							& $11.499 \pm 0.043$				& \textit{Gaia} DR2\\
$T_\mathrm{eff}$ [K]						& $4403^{+46}_{-53}$				& This work\\
log $g$ [dex]							& $4.500^{+0.011}_{-0.012}$			& This work\\
$V$ [mag]							& 12.4						& E2017\\
$G$ [mag]							& 11.9						& \textit{Gaia} DR2\\
$M_\mathrm{s}$ [$M_\odot$]					& $0.688 \pm 0.037$				& E2017\\
$R_\mathrm{s}$ [$R_\odot$]					& $0.6506 \pm 0.0054$				& E2017\\
$K$ [$\mathrm{m.s}^{-1}$]					& $551.0 \pm 3.2$				& E2017\\
\hline
$T_\mathrm{c}$ [$\mathrm{BJD}_\mathrm{TDB}$]			& $2~455~528.869175^{+0.000061}_{-0.000063}$	& This work\\
$P_\mathrm{orb}$ [d]						& $0.813473949^{+0.000000019}_{-0.000000021}$	& This work\\
$i$ [deg]							& $81.684^{+0.039}_{-0.033}$ 			& This work\\
$R_\mathrm{s}/a$ 						& $0.21127^{+0.00056}_{-0.00063}$		& This work\\
$R_\mathrm{p}/R_\mathrm{s}$ (\textit{TESS} passband)		& $0.16015^{+0.00040}_{-0.00042}$		& This work\\
$R_\mathrm{p}/R_\mathrm{s}$ ($g$ passband)			& $0.16630^{+0.00120}_{-0.00110}$		& This work\\
$R_\mathrm{p}/R_\mathrm{s}$ ($r$ passband) 			& $0.16412^{+0.00051}_{-0.00064}$		& This work\\ 
$R_\mathrm{p}/R_\mathrm{s}$ ($i$ passband)			& $0.16740^{+0.00120}_{-0.00100}$		& This work\\
$R_\mathrm{p}/R_\mathrm{s}$ ($z_\mathrm{s}$ passband)		& $0.16530^{+0.00150}_{-0.00180}$		& This work\\
$R_\mathrm{p}/R_\mathrm{s}$ (all passbands)			& $0.16224 \pm 0.00031$				& This work\\
$R_\mathrm{p}$ [$\mathrm{R}_\mathrm{Jup}$] 			& $1.037^{+0.022}_{-0.019}$			& This work\\
$M_\mathrm{p}$ [$\mathrm{M}_\mathrm{Jup}$]			& $1.997^{+0.072}_{-0.073}$			& This work\\
$\rho_\mathrm{p}$ [$\mathrm{g.cm}^{-3}$]			& $2.10^{+0.31}_{-0.32}$	 		& This work\\
$a$ [a.u.]							& $0.01507^{+0.00026}_{-0.00027}$		& This work\\
\hline
\hline
\end{tabular}
\end{table}

Based on the joint \textit{TESS} and MuSCAT2 observations we obtained an orbital period of $P_\mathrm{orb} = 0.813473949^{+0.000000019}_{-0.000000021}$ d and $T_\mathrm{c} = 2~455~528.869175^{+0.000061}_{-0.000063}$ $\mathrm{BJD}_\mathrm{TDB}$. We refined this ephemeris in Sect. \ref{trantimings} based on the whole O-C data-set of mid-transit times, therefore this result is considered as a preliminary orbital period and reference mid-transit time. We note, however, that this ephemeris was also necessary during the timing analysis -- the O-C diagram of the mid-transit times (see Fig. \ref{ocdiagram}) was calculated based on this $P_\mathrm{orb}$ and $T_\mathrm{c}$, using the Eq. \ref{lineph}. The orbit inclination angle value, $i = 81.684^{+0.039}_{-0.033}$ deg, obtained by the \rmf code, is in a $3\sigma$-agreement with the parameter values presented, e.g., by \citet{Hellier1} and by \citet{Ricci1}, i.e., $i = 82.6^{+1.3}_{-0.9}$ deg and $i = 81.92 \pm 0.54$ deg, respectively. \citet{Hoyer1} obtained $i =  82.11 \pm 0.10$ deg, which is out of this $3\sigma$-agreement interval. We found the value of $R_\mathrm{s}/a = 0.21127^{+0.00056}_{-0.00063}$, which is equal to $a/R_\mathrm{s} = 4.733^{+0.012}_{-0.014}$. This parameter value corresponds within $3\sigma$ to the parameter values presented, e.g., by \citet{Ricci1} and by \citet{Esposito1}, i.e., $a/R_\mathrm{s} = 4.82 \pm 0.11$ and $a/R_\mathrm{s} = 4.97 \pm 0.14$, respectively. \citet{Hoyer1} obtained $a/R_\mathrm{s} = 4.867 \pm 0.023$, which is about $5.8\sigma$ difference.

In the case of the planet-to-star radius ratio parameter, we obtained the best-fitting value of \rprs = $0.16224 \pm 0.00031$ for all passbands combined, which is comparable, e.g., with the value of \rprs = $0.1588 \pm 0.0040$, published by \citet{Esposito1}. Several transmission spectroscopy observations targeted also WASP-43, see e.g., \citet{Chen1}, \citet{Kreidberg3}, \citet{Murgas1}, or \citet{Ricci1}. The passband-dependent planet-to-star radius ratio parameter values derived during the joint analysis are in a $3\sigma$ agreement with these transmission spectra. Mainly the \textit{TESS} passband value fit well the existing measurements. 

\subsection{Transit timings and ephemeris refinement}
\label{trantimings}

\begin{figure*}
\centering
\centerline{
\includegraphics[width=\columnwidth]{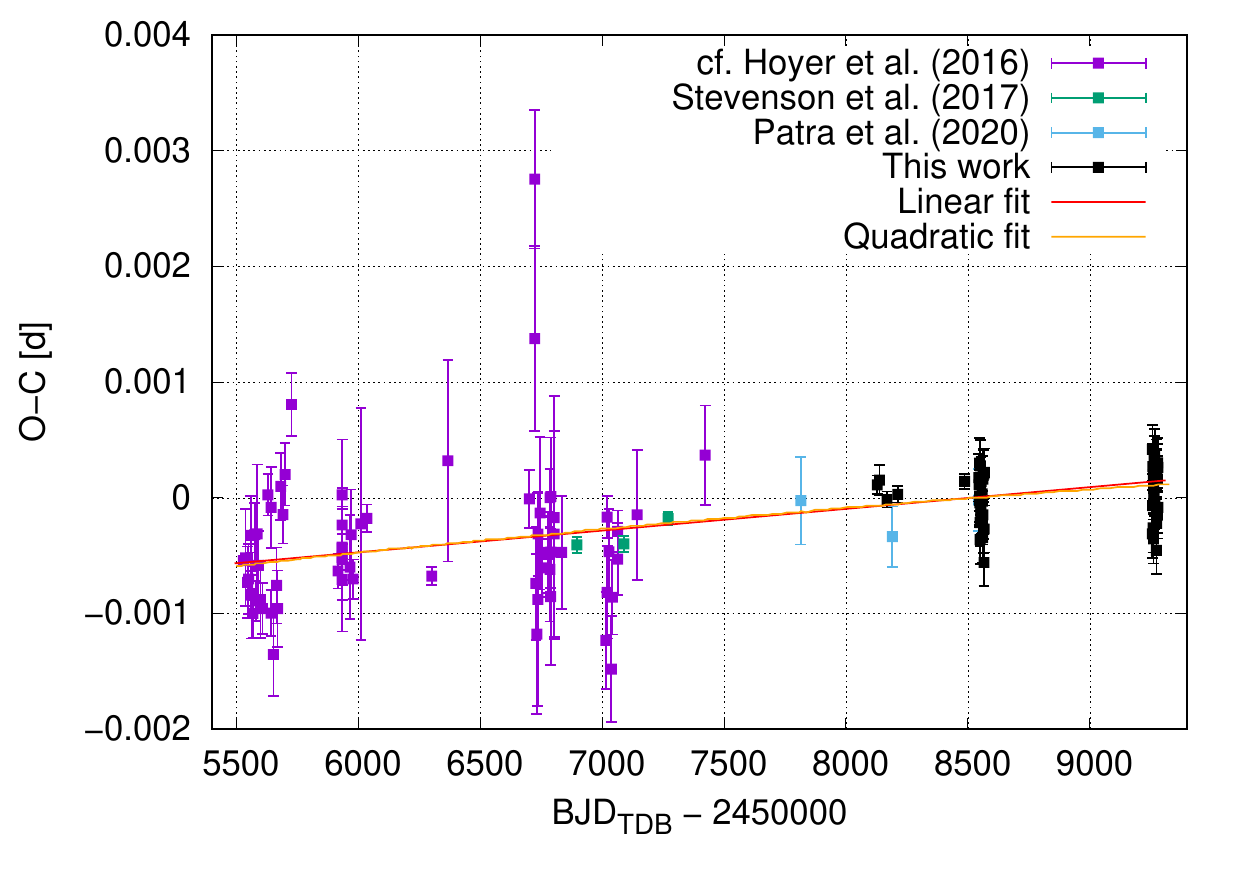}
\includegraphics[width=\columnwidth]{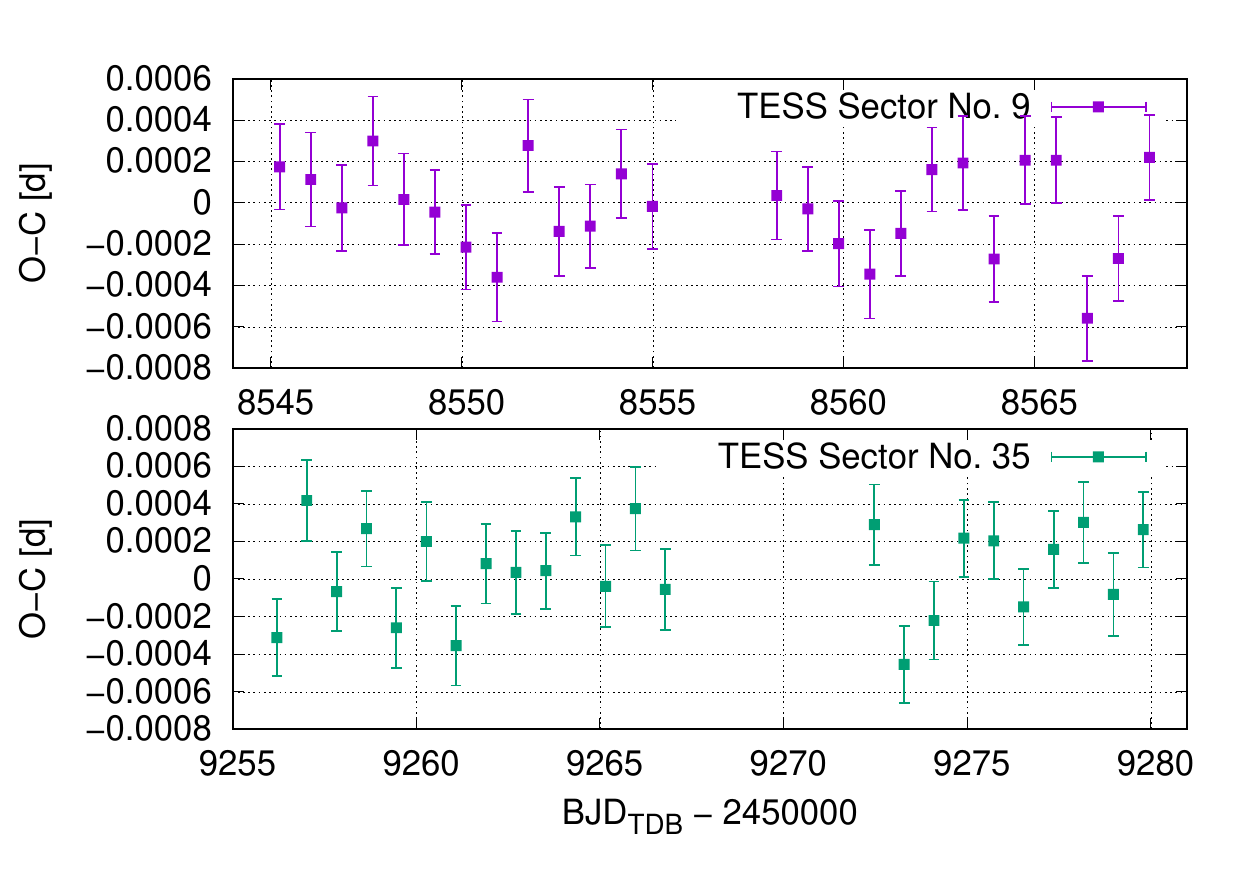}}
\caption{Observed-minus-calculated (O-C) diagrams of WASP-43b mid-transit times. The left-hand panel shows the whole O-C data-set sorted by the sources and fitted with a linear/quadratic function using the {\tt{OCFIT}} code. The right-hand panel focuses on the O-C data-points derived based on the \textit{TESS} data for better visibility.}
\label{ocdiagram} 
\end{figure*} 

The whole O-C data-set of mid-transit times used during our analysis is presented in Fig. \ref{ocdiagram} (left-hand panel). In this graph we can see the 68 data-points, collected by \citet{Hoyer1}, the 3 data-points, observed by \citet{Stevenson1}, the 3 data-points presented by \citet{Patra2}, the 50 data-points derived from the \textit{TESS} observations, plotted also separately in Fig. \ref{ocdiagram} (right-hand panel), and the 5 data-points derived from the MuSCAT2 observations in this work, i.e., 129 data-points in total. As first, these O-C data were fitted with a linear function. This fit is shown in Fig. \ref{ocdiagram} (left-hand panel). Based on the linear fit we obtained a refined linear ephemeris of:

\begin{equation}
\label{ttvlinfit}
T_0 = 2~455~528.868607(51)~\mathrm{BJD}_\mathrm{TDB} + 0.813474101(18)~\mathrm{d} \times E.
\end{equation}      

\noindent Subsequently, the O-C data-set of mid-transit times was fitted with a quadratic function. This fit is also depicted in Fig. \ref{ocdiagram} (right-hand panel). We can see that the quadratic trend is not significant at the first glance. Based on this quadratic fit we obtained a quadratic ephemeris of:

\begin{equation}
\label{}
T_0 = 2~455~528.868590(60)~\mathrm{BJD}_\mathrm{TDB} + 0.813474135(70)~\mathrm{d} \times E +  Q \times E^2, 
\end{equation}    
  
\noindent where the quadratic coefficient is $Q = (-0.8 \pm 1.5) \times 10^{-11}$ d, confirming the negligible quadratic trend in the O-C data-set of mid-transit times. Based on the Eq. \ref{qqq} we can easily calculate the period change rate $\dot{P}$ from $Q$. We obtained a dimensionless value of $\dot{P} = (-2.0 \pm 3.7) \times 10^{-11}$, which we can convert to $\dot{P} = -0.6 \pm 1.2$ ms~yr$^{-1}$. The result is negative, but is not significant, confirming the previous results about no detection. In comparison with the results presented by \citet{Hoyer1}, i.e., $\dot{P} = 0.0 \pm 6.6$ ms~yr$^{-1}$, derived by \citet{Stevenson1}, i.e., $\dot{P} = +9.0 \pm 4.0$ ms~yr$^{-1}$, and obtained by \citet{Patra2}, i.e., $\dot{P} = +14.4 \pm 4.6$ ms~yr$^{-1}$, our result is more precise thanks to the high quality \textit{TESS} observations and to the longer time baseline. The quality of the linear and quadratic fit was expressed as Bayesian Information Criterion ($BIC$), which is defined as:

\begin{equation}
BIC = \chi^2 + k \ln N,
\label{bic}
\end{equation}

\noindent where $k$ is the number of free parameters of the model and $N$ is the number of data-points. There is no significant difference between the two Bayesian Information Criterions, i.e., $BIC = 266.1$ in the case of the linear fit and $BIC = 269.9$ in the case of the quadratic fit, which means that it is not justified to use the quadratic fit. 

Since we did not detect significant orbital period change rate of WASP-43b, we can adopt the values presented in Eq. \ref{ttvlinfit} as a final solution for the reference mid-transit time and orbital period of the planet, i.e., $T_\mathrm{c} = 2~455~528.868607 \pm 0.000051$ $\mathrm{BJD}_\mathrm{TDB}$ and $P_\mathrm{orb} = 0.813474101 \pm 0.000000018$ d. In comparison with the ephemeris obtained from the joint \textit{TESS} and MuSCAT2 analysis (see Table \ref{wasp43outputs}), using the whole O-C data-set of mid-transit times we improved the parameter $T_\mathrm{c}$ by a factor of 1.23 and the parameter $P_\mathrm{orb}$ by a factor of 1.16. This result confirms the previously published parameter values, see Table \ref{wasp43eph} for examples.          

\begin{table}
\centering
\caption{The finally adopted ephemeris of WASP-43b, improved based on the whole O-C data-set of mid-transit times, and its comparison with two previously published ephemeris. Notes: H2011 = \citet{Hellier1}, H2016 = \citet{Hoyer1}.}
\label{wasp43eph}
\begin{tabular}{lcc}
\hline
\hline
Parameter & Value & Source\\
\hline
\hline
$T_\mathrm{c}$ [$\mathrm{BJD}_\mathrm{TDB}$] & $2~455~528.868607 \pm 0.000051$ & This work\\  
$P_\mathrm{orb}$ [d] & $0.813474101 \pm 0.000000018$ & This work\\
\hline
$T_\mathrm{c}$ [$\mathrm{BJD}_\mathrm{TDB}$] & $2~455~528.86774 \pm 0.00014$ & H2011\\
$P_\mathrm{orb}$ [d] & $0.8134750 \pm 0.0000010$ & H2011\\
\hline
$T_\mathrm{c}$ [$\mathrm{BJD}_\mathrm{TDB}$] & $2~455~528.868634 \pm 0.000046$ & H2016\\
$P_\mathrm{orb}$ [d] & $0.813473978 \pm 0.000000035$ & H2016\\
\hline 
\hline
\end{tabular}
\end{table}

\section{Conclusions}
\label{concl}         

Using the O-C data-set of mid-transit times, derived from 129 transits of WASP-43b, we have re-estimated the orbital period change rate of the exoplanet. The obtained result, $\dot{P} = -0.6 \pm 1.2$ ms~yr$^{-1}$, is consistent with a constant period well within $1\sigma$. It confirms the previous results about no detection, but in comparison with the previous results, our result is more precise thanks to the high quality \textit{TESS} observations and to the longer time baseline. By extending the observations to more than 730 days, i.e., covering a time baseline of about 10 years, we could improve the $\dot{P}$ parameter by a factor of about 3.8 in comparison with the latest published result. Thus, we see no evidence to support previous claims of a decaying orbit for WASP-43b. As a by-product of the data analysis, we also derived the system parameters of WASP-43b. Several of them were refined in comparison with the previously published parameters. Thanks to combination of the high quality \textit{TESS} and multi-color MuSCAT2 observations we could estimate, for example, the orbit inclination angle parameter and the combined planet-to-star radius ratio parameter precisely and without parameter degeneration. The derived parameter values are mostly in $3\sigma$ agreement with the results of existing studies. Since we did not detect significant period change rate of the planet, a new linear ephemeris of WASP-43b was derived using the whole O-C data-set of mid-transit times.
 

\section*{Acknowledgements}

We thank Dr. S. Hoyer from the Laboratoire d'Astrophysique de Marseille (LAM) in France for the helpful discussions. We also thank the anonymous reviewer for the helpful comments and
suggestions. This work was supported by the ERASMUS+ grant No. 2017-1-CZ01-KA203-035562, by the VEGA grant of the Slovak Academy of Sciences No. 2/0031/18, by an ESA PRODEX grant under contracting with the ELTE University, by the GINOP No. 2.3.2-15-2016-00003 of the Hungarian National Research, Development and Innovation Office, and by the City of Szombathely under agreement No. 67.177-21/2016. This article is based on observations made with the MuSCAT2 instrument, developed by ABC, at Telescopio Carlos S\'{a}nchez operated on the island of Tenerife by the IAC in the Spanish Observatorio del Teide. This work is partly financed by the Spanish Ministry of Economics and Competitiveness through grants No. PGC2018-098153-B-C31. This work is partly supported by JSPS KAKENHI grant Nos. JP18H01265 and JP18H05439, and JST PRESTO grant No. JPMJPR1775. This work is partly supported by JSPS KAKENHI grant No. JP17H04574. This work was partly supported by Grant-in-Aid for JSPS Fellows, grant No. JP20J21872. M.T. is supported by MEXT/JSPS KAKENHI grant Nos. 18H05442, 15H02063, and 22000005. A.C. acknowledges financial support from the State Agency for Research of the Spanish MCIU through the ''Center of Excellence Severo Ochoa'' award for the Instituto de Astrophysics of Andalusia (SEV-2017-0709). We acknowledge funding from the European Research Council under the European Union's Horizon 2020 research and innovation program under grant agreement No 694513.

\section*{Data availability}

The data underlying this article will be shared on reasonable request to the corresponding author. The reduced light curves presented in this work will be made available at the
CDS (\url{http://cdsarc.u-strasbg.fr/}).




\bibliographystyle{mnras}
\bibliography{Yourfile} 






\bsp	
\label{lastpage}
\end{document}